\documentclass{mn2e}

\newcommand{\msunyr}{\mbox{\,{\rm M}$_\odot$\ {\rm yr}$^{-1}$\,}}
\newcommand{\msun}{\mbox{\,{\rm M}$_\odot$\,}}
\newcommand{\kms}{\mbox{${\,\rm km~s}^{-1}$}\,}
\newcommand{\ergs}{\mbox{${\,\rm erg~s}^{-1}$}\,}
\newcommand{\etal}{{et al.}}

\begin{document}

\title[Magnetospheric Radio Emission from Extrasolar Giant Planets] 
{Magnetospheric Radio Emission from Extrasolar Giant Planets: The Role
of the Host Stars} 
 
\author[Ian R. Stevens]
{Ian R. Stevens\\ School of Physics and Astronomy, 
University of Birmingham, Edgbaston, Birmingham B15 2TT \\
(E-mail: irs@star.sr.bham.ac.uk)}

\maketitle

\begin{abstract}

We present a new analysis of the expected magnetospheric radio emission
from extrasolar giant planets for a distance limited sample of the
nearest known extrasolar planets. Using recent results on the
correlation between stellar X-ray flux and mass-loss rates from nearby
stars, we estimate the expected mass-loss rates of the host stars of
extrasolar planets that lie within 20~pc of the Earth. We find that some
of the host stars have mass-loss rates that are more than $100$ times
that of the Sun, and given the expected dependence of the planetary
magnetospheric radio flux on stellar wind properties this has a very
substantial effect.  Using these results and extrapolations of the
likely magnetic properties of the extrasolar planets we infer their
likely radio properties.

We compile a list of the most promising radio targets, and conclude
that the planets orbiting Tau Bootes, Gliese~86, Upsilon Andromeda and
HD\,1237 (as well as HD\,179949) are the most promising candidates,
with expected flux levels that should be detectable in the near future
with upcoming telescope arrays. The expected emission peak from these
candidate radio emitting planets is typically $\sim 40-50$~MHz. We
also discuss a range of observational considerations for detecting
extrasolar giant planets.

\end{abstract}


\section{Introduction} 
\label{sec1}

The reasons for searching for and studying extrasolar giant planets
(EGPs) at radio wavelengths are many. First, to provide another means
of direct detection of EGPs, second, if a detection with sufficient
strength is made, it will provide evidence of the presence and indeed
strength of the magnetic field of the EGP. Many current planet-finding
techniques are indirect in nature, relying on seeing the effect of the
planet on the host star (such as the Doppler, transit, astrometric or
microlensing methods). In the future, direct detection of the planet
may be made using space-based interferometric techniques, or through
coronographic techniques. Radio detection provides another possible
means of directly observing the planet (or the planet's near
environment).

In principle, the detection
of radio emission could also provide information on the existence of any
satellites around the EGP, the rotation period, and indeed provide
constraints on the inclination of the EGP (though these last reasons are
very speculative and will require high quality observations). For the
time being, mere detection of EGPs at radio wavelengths is a
sufficiently worthwhile goal.

The solar system object that leads to the expectation that EGPs may be
detectable at radio wavelengths is Jupiter (and, to a lesser extent, the
other planets that have been detected at radio wavelengths, namely the
Earth and the other Jovian planets -- see Bastian, Dulk \& Leblanc 2000). At
decameter wavelengths, Jupiter is an extremely bright source, detectable
with very modest equipment. At these wavelengths the Sun/Jupiter
contrast can be $\sim 1$ and indeed the polarised nature of the emission
from Jupiter (and we expect from EGPs) will allow discrimination as to
whether the emission is planetary or stellar in origin. We shall discuss 
the Jovian radio emission in Section~\ref{sec1p1}.

The radio emission from the solar system planets is related to their
magnetospheres, and a radio detection of an EGP will give direct
evidence of a magnetosphere. However, indirect evidence for a planetary
magnetosphere has already come to light in the case of HD\,179949\,b
(Shkolnik, Walker \& Bohlender 2003), where variations in the Ca{\small
II} H and K lines seem to be phase locked with the orbit of the planet
(which has a period of 3.093~days and a mass of $M_p\sin i =
0.98M_{Jup}$). A number of other stars with short period EGPs were also
studied in the same way with no other comparable results.  HD~179949 is
more distant than the sample covered in this paper (at a distance of
$D=27$~pc), but because of the inferred existence of a magnetosphere we
will include it in the sample discussed below.

\subsection{Origin of the Jovian Radio Emission}
\label{sec1p1}

In the magnetospheres of planets, intense radio emission can result
from electron cyclotron maser radiation. In this process keV electrons
in the auroral regions of the planet radiate at the gyro-frequency of
the magnetic field lines. The radiation is emitted in a hollow cone
region, and is highly structured in time and frequency.  The
requirements for this process are simply a magnetic field and a source
of energetic (keV) electrons. The presence of a magnetic field will be
extremely likely in EGPs, and we would expect the magnetic moment of
the more massive EGPs to exceed that of Jupiter
(Section~\ref{sec2p4p1}). The energetic electrons can be provided by a
range of means (in the case of solar system objects); the solar wind,
auroral processes or planet-moon magnetic coupling (in the case of
Io/Jupiter) and there is no reason to expect this to be different for
EGPs.  The electron-cyclotron maser emission is seen to be very
sporadic in the case of Jupiter and the other planets, and the
emission is 100\% circularly or elliptically polarised. For a
discussion of the extremely variable S-burst decametric emission and
the rather more slowly varying L-type emission, see Queinnec \& Zarka
(2001).

In the case of Jupiter, the peak emission occurs at a frequency of $\sim
10-20$~MHz (though it is seen to extend up to $\sim 40$~MHz), while for
the other magnetised planets it is at correspondingly lower frequencies,
in the range of $0.1-1$~MHz. For planets other than Jupiter, this
emission is not observable from the ground (with the ionospheric cut-off
typically in the range of $\sim 2-15$~MHz, depending on the location on
Earth, day or night, location in the solar cycle etc), and were detected
by space-based instruments. Jupiter has also been detected as a radio
source at higher (GHz) frequencies, but at a much lower level.

One further constraint, is that the local plasma frequency in the source
region must be lower than the gyro-frequency, which is the case for most
solar system objects. The exception to this is probably Mercury, where
the local plasma frequency in the solar wind exceeds the
gyro-frequency (the surface magnetic field at the equator is
$B=0.0033$~G). As discussed in Section~\ref{sec3}, there will also
be issues with this associated with planets with stronger magnetic
fields immersed in more dense stellar winds, or orbiting at very close
distances to the host star.

For an overview of the emission mechanisms and the radio properties of the
solar system planets see Bastian \etal\ (2000), and references therein.

\subsection{Previous Searches for EGP Radio emission}
\label{sec1p2}

There have been a small number of reported searches (or related
proposals for searches) for radio emission from extrasolar planets, with
no detection as yet. The searches reported include:

\begin{enumerate} \item Winglee, Dulk \& Bastian (1986) used the Very
Large Array ({\it VLA}) to observe six nearby stars at 333 and
1400~MHz. These objects are not part of the current list of EGP hosts
(with the possible exception of Lalande~21185). None were detected.

\item Bastian \etal\ (2000) reported on {\it VLA} observations of a range of
EGP and brown dwarf candidates at 333~MHz and 1465~MHz, with a smaller
number of observations at 74~MHz. The relevant objects covered included
51~Peg, Ups~And, 55~Cnc, 47~UMa, Tau~Boo, 70~Vir and
Lalande~21185. These object were all observed at 333~MHz and 1465~MHz,
while 47~UMa was also observed at 74~MHz. Of these, the 74~MHz
observations are the most relevant and the {\sl rms} noise level for the
non-detection the 47~UMa observation was quoted as 76~mJy for the
entire observing run and 3.24~Jy for a 10~sec period (important if the
emission is extremely variable, which if Jupiter is anything to go by,
is likely to be the case).

\item Farrell \etal\ (2003) observed Tau~Boo with the {\it VLA} at 74~MHz,
with a quoted upper flux limit of 0.12~Jy (see also Farrell, Desch \&
Zarka 1999 for an earlier analysis of the likely radio emission from
Tau~Boo).

\item Ryabov, Zarka \& Ryabov (2003) reported on progress on
observations of EGPs, using the UTR-2 array in the Ukraine. Observations
of 20 nearby EGPs have been undertaken with no detections reported so
far.

\item While not a search as such, Butler (2003) reports on the prospect
of radio detection of EGPs with upcoming instruments such as the Square
Kilometer Array ({\it SKA}) or the Low-Frequency Array ({\it LOFAR}
\footnote{For a description of the proposed {\it LOFAR} instrument see
http://www.lofar.org}).

\end{enumerate}

We shall discuss these observational limits on the radio emission from
EGPs, in the light of the new predictions presented here, in
Section~\ref{sec4}.

The rest of the paper is set out as follows: in Section~2, the basic
theory of radio emission from EGPs is introduced, including a discussion
of the influence of the planetary parameters and characteristics of the
stellar winds of the host star on magnetospheric radio emission from
EGPs; in Section~3, the results from the modelling are presented, and in
Section~4, these results are discussed, along with comments on the
likely observability of the radio emission from EGPs.

\section{Radio Emission from EGPs}
\label{sec2}

\subsection{The Solar System: Physics And Extrapolations}
\label{sec2p1}

As noted by Zarka \etal\ (2001), following on from work by Desch \&
Kaiser (1984) and Zarka (1992), the level of planetary radio emission is 
proportional to the kinetic flux intercepted by the planetary
magnetosphere, and, as a consequence of the frozen-in field condition,
also proportional to the incident magnetic energy flux. Bastian \etal\ (2000)
have also reported on results based on this relationship, termed the
``radiometric Bode's Law'' for the solar system.

If the incident solar wind kinetic flux on the planetary
magnetosphere is $P_{ram}$ and the incident magnetic energy flux on the
magnetosphere is $P_{mag}$, and the emitted radio power from the planet
is $P_r$, then this relationship states that:

\begin{equation}
P_r=\alpha P_{ram}
\end{equation}
with $\alpha$ an efficiency parameter, with $\alpha\sim 7\times
10^{-6}$, and 
\begin{equation}
P_r=\beta P_{mag}
\end{equation}
with $\beta$ another efficiency parameter, with $\beta=3\times
10^{-3}=400\alpha$.

The similar form of these relationships is a consequence of
the azimuthal component of the magnetic field in the solar wind scaling
as $B_\phi\propto r^{-1}$ where $r$ is the distance from the
star, and so that the magnetic energy flux scales as $r^{-2}$. The
kinetic flux scales as $\rho V_w^3$ (with $\rho$ the wind density and
$V_w$ the wind velocity). For a spherical and constant velocity wind
then this scales as $r^{-2}$ as well.

As noted by Zarka \etal\ (2001), the high apparent efficiency of
conversion of magnetic energy density into radio power might suggest
that the conversion of incident magnetic energy, via reconnection, into
particle acceleration (resulting in particles following the magnetic
field lines towards the planet) is the process that actually leads to
the radio emission.

Based on this relationship we will generalise the expected radio power
from solar system planets for arbitrary planetary and stellar parameters.

\subsection{The Sample of EGPs}
\label{sec2p2}

In order to investigate the expected properties of known EGPs we shall
concentrate on the nearest EGPs, where we will have the greatest
possibility of detection, and where the X-ray fluxes for a reasonable
fraction of the host stars are known. The NEXXUS database
\footnote{The NEXXUS database is available at
http://www.hs.uni-hamburg.de/DE/For/Gal/Xgroup/nexxus} 
is the most comprehensive database of X-ray data
on nearby stars and contains data on stars out to a distance of 25~pc (see
Section~\ref{sec2p5p1} for a fuller discussion of the use of this database).
We will concentrate on those EGPs that lie within 20~pc, and provide
estimates of the relevant stellar and planetary parameters and from
these infer likely radio properties. In Table~1 and Table~2 we list the
details of the host stars and the EGPs that lie within 20~pc of the
Earth, and in the following sections we shall describe the derivation of
the values in these tables. We shall also include results for
HD\,179949, which although lying outside of 20~pc, has been inferred to
have a magnetosphere via other means (Shkolnik \etal\ 2003).

\begin{figure*}
\vspace*{8cm}
\includegraphics{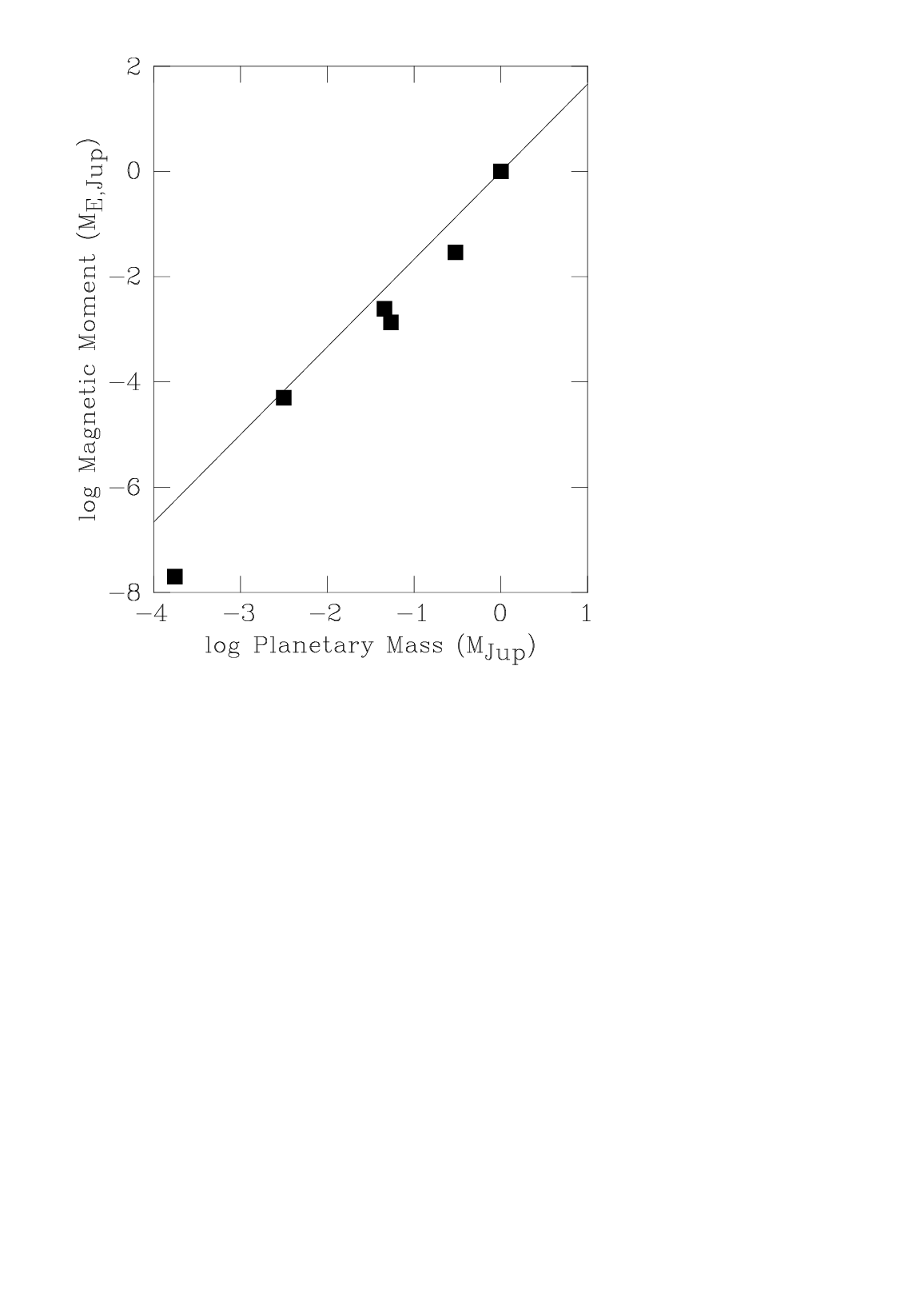} 
\includegraphics{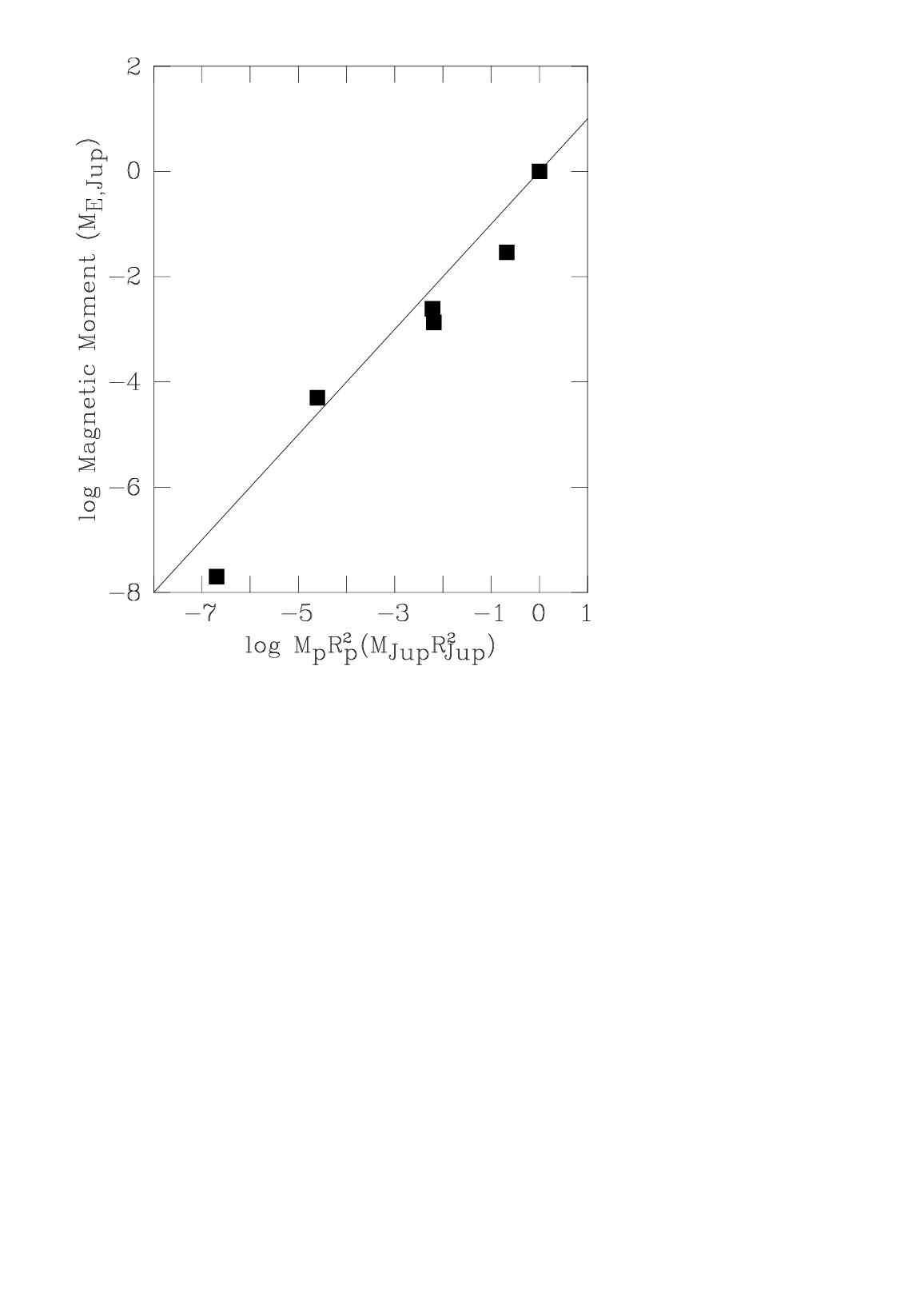} 
\caption{Left panel: the planetary magnetic moment $M_E$ (in
units of the value for Jupiter, $M_{E,Jup}$) for the six magnetised 
solar system planets (in terms of increasing mass, 
Mercury, Earth, Uranus, Neptune, Saturn and Jupiter) versus the 
planetary mass $M_p$ (in units of the mass of Jupiter $M_{Jup}$). 
The plotted line is not a fit, just a schematic relationship with 
$M_E\propto M_p^{1.66}$, and plotted so that it passes through the point 
for Jupiter.
Right panel: the same as for the left panel, except this time the
planetary magnetic moment $M_E$ is plotted against $M_p R_p^2$ for the
solar system planets, with $M_p$ and $R_p$ the planetary mass and radius
respectively. The line plotted here is again a schematic plot, with
$M_E\propto (M_p R_p^2)^{1.0}$, with the line plotted so that it passes
through the value for Jupiter. See Section~\ref{sec2p4p1} for a fuller
discussion of these relationships.}
\label{fig1}
\end{figure*}

\subsection{General Magnetospheric Emission from EGPs}
\label{sec2p3}

In order to have a description of the expected radio power from the
magnetospheres of exoplanets we need to describe the properties of the
host star (stellar wind properties such as mass-loss rate and velocity
and magnetic field strength), and the properties of the exoplanets
(planetary mass, radius, magnetic moment and orbital parameters).

We also need to introduce two important frequencies that will enter
into the discussion of the expected radio emission. The first is the
characteristic frequency of a plasma (referred to as the plasma frequency):

\begin{equation} 
\nu_{pe} ({\rm MHz})=\left(\frac{n_e e^2}{\pi m_e}\right)^{1/2}=
8.98\times 10^{-3} n_e^{1/2} 
\end{equation} 
with the electron density $n_e$ in cm$^{-3}$. The plasma frequency
provides a lower frequency limit for the propogation of electromagnetic
radiation; that is radiation with a frequency lower than the local
plasma frequency will be screened out by the plasma.

The second is the electron
gyro-frequency (or electron cyclotron frequency):
\begin{equation}
\nu_{ce} ({\rm MHz})=\left(\frac{eB}{2\pi m_ec}\right) = 2.80 B 
\end{equation}
with the magnetic field strength $B$ in Gauss.
The gyro-frequency is important as it relates to the frequency of the
peak radio emission (see Section~\ref{sec2p4}). The plasma frequency
will also be important as it sets a lower limit on the frequency,
whereby waves with a frequency below the plasma frequency will be
screened out. This will be important in the context of EGPs in very
close orbits or with stars with high mass-loss rates.

For the basic stellar properties of the host star, we assume that we
have a star of mass $M_\ast$, radius $R_\ast$, effective temperature
$T_\ast$ and equatorial magnetic field $B_\ast$, which is located at a
distance $D$ from the Earth.

We shall consider the situation in a very general manner, though in
practice we shall mostly concentrate on the stellar types of stars with
known EGPs, namely main-sequence F, G, K and M-stars, which will have
coronally driven winds, similar to that of the Sun.

We shall assume that the stellar wind of the star has a mass-loss rate
of $\dot M_\ast$ and a wind velocity of $V_w$ (in the case of
short-period planets the wind velocity will not be the terminal
velocity, see Section~\ref{sec3}). Of course, mass-loss from
stars tends not to be isotropic, and this is certainly true of the Sun
and probably true of most stars. However, for simplicity (and for the
order of magnitude arguments being employed here) we shall stick with
spherically symmetric winds.

Except in the inner regions of the heliosphere, the magnetic field is
dominated by the azimuthal component, and so that if the surface magnetic
field is $B_\ast$, stellar radius is $R_\ast$, and the angular frequency
of the stellar rotation is $\Omega_\ast$, then the magnetic field at a
radius $r$ is given by

\begin{equation}
B_\phi(r)=B_\ast\left(\frac{R_\ast^2 \Omega_\ast}{V_w r}\right)
\end{equation}
and so that the magnetic energy ($B^2/8\pi$) scales as $r^{-2}$. This
will not be the case in the inner wind region, where the radial
component will begin to dominate, with

\begin{equation}
B_r(r)=B_\ast\left(\frac{R_\ast}{r}\right)^2 \, .
\end{equation}

The radius at which the azimuthal component starts to dominate depends
on the stellar rotation rate (as well as things like the wind
velocity), and some of the more interesting systems in the sample are
rapid rotators (this is because they are interesting from a radio
point of view due to high X-ray activity, which in turn is related to
youth and rapid rotation). For instance, Barnes (2001) quotes a
rotation period of 3.2 days for Tau Boo. This means that in these
cases the azimuthal magnetic field will start to dominate at
correspondingly smaller radii.

As discussed by Zarka \etal\ (2001), the inclusion of the radial field
component makes the magnetic field strength rise more sharply as the
radius decreases, than if we just included the azimuthal field, which in 
turn should make the anticipated radio flux rise more steeply (though
with some limited exceptions -- see Zarka \etal\ 2001). In this paper we 
shall just include the azimuthal field in our estimations. This
assumption is appropriate for the longer period planets and should be on 
the conservative side for the shorter period planets.

We assume that the star is orbited by an EGP. The planet is assumed to
orbit with a period $P_{orb}$, at a distance $A$ from the star. We shall
assume that all planets have a circular orbit for the time being, so
that $A$ is equal to the orbital semi-major axis of the planet (which we
denote as $a$). This is clearly a poor assumption given that most EGPs
have eccentric orbits, with the eccentricities extending up to
$e=0.7$. We will discuss the impact of orbital eccentricity in
Section~\ref{sec3p2}. The planet is assumed to have a mass $M_p$, radius
$R_p$ and magnetic moment $M_E$.

From the radiometric Bode's law (and following Zarka \etal\ 2001) we
have that the emitted radio power is proportional to both the magnetic
energy and kinetic flux incident on the magnetosphere, so that

\begin{equation}
P_r=\beta P_{mag}= \beta \epsilon_{mag} V_w \pi R_{MP}^2
\end{equation}
with $\epsilon_{mag}=B(r)^2/8\pi$ and $R_{MP}$ the magnetospheric radius
of the planet. This equation can also be written as 
\begin{equation}
P_r=\alpha P_{ram}= \alpha \epsilon_{ram} V_w \pi R_{MP}^2
\end{equation}
with $\epsilon_{ram}=(\rho V_w^2)$, with $\rho$ the stellar wind
density, with
\begin{equation}
\rho=\frac{\dot M_\ast}{4 \pi r^2 V_w}
\end{equation}
with $\dot M_\ast$ the stellar mass-loss rate.

We are beginning to have some knowledge of the mass-loss rates of the
host stars of EGPs, with the X-ray surface flux being an indicator,
and so we have an indication of the stellar wind ram-pressure
(Section~\ref{sec2p5}). On the other hand we have little knowledge of
the surface magnetic fields of these stars. Consequently, we shall use
the form of the radiometric Bode's law, involving the incident kinetic
flux, with the same value of $\alpha$ as for the solar system.

We note that we are effectively assuming that $\beta/\alpha=400$, and
this will be universal for stars other than our Sun. This assumption 
leads to the following relationship, that 

\begin{equation}
\frac{\beta}{\alpha}=\frac{2\dot M_\ast V_w^3}{B_\ast^2 R_\ast^4
\Omega_\ast^2}\ , 
\end{equation}
so that an increase in the stellar mass-loss rate of a factor $x$
corresponds to an increase in stellar surface magnetic field of
$x^{1/2}$ (all other things being equal). 

Returning to the overall picture, the planetary magnetospheric
radius ($R_{MP}$) will be a function both of the planetary magnetic
moment $M_E$ and the stellar wind characteristics, with

\begin{equation}
R_{MP}=\left[\frac{C M_E^2}{16\pi\rho V_w^2}\right]^{1/6}\propto
M_E^{1/3}{\dot M_\ast}^{-1/6} V_w^{-1/6} A^{1/3}
\end{equation}
where $\rho$ and $V_w$ are the density and velocity of the stellar wind
at the radius of the planet, and $C$ is a constant. 
We determine the magnetospheric radii of
the EGPs by assuming that the magnetospheric radius of Jupiter is
$45R_{Jup}$, and scale accordingly according to stellar mass-loss rate
and planetary distance from the star (and assuming $\dot M_\odot=2\times 
10^{-14}\msunyr$ and $V_w=400\kms$).

The kinetic flux of the stellar wind intercepted by the
magnetosphere of the planet, orbiting at a radius $A$, can be written as

\begin{equation}
P_{ram}=\rho V_w^3 \pi R_{MP}^2 =\frac{\dot M_\ast V_w^2 R_{MP}^2}{4 A^2 }
\end{equation}
and consequently the expected emitted radio flux from the planet ($P_r$) 
will be
\begin{equation}
P_{r}=\alpha \frac{\dot M_\ast V_w^2 R_{MP}^2}{4 A^2 } \ .
\end{equation}
Substituting in for $R_{MP}$, this leads to the following
scaling relationship,  
\begin{equation}
P_{r} \propto {\dot M}_\ast^{2/3} V_w^{5/3} M_E^{2/3} A^{-4/3}
\end{equation}
So, for example, although moving a planet closer means that the stellar wind
density scales as $A^{-2}$, the planetary magnetosphere size reduces as
well, partially offsetting this increase. The same process occurs for
increasing the mass-loss rate and so on.
The flux detected at Earth will obviously have an additional $D^{-2}$
dependency. 

For Jupiter, we shall assume a peak in the emission at 10~MHz, and a flux
density of $10^8$~Jy at an equivalent distance of 1~AU (Bastian \etal\
2000). Based on this, for each EGP we can estimate the expected radio
flux and peak frequency for this flux. The planetary parameters required
to estimate these values are discussed below. In estimating the
expected flux from these EGPs we obviously have to include the distance $D$
of the host star from the Earth.

\subsection{Planetary Parameters}
\label{sec2p4}

As discussed earlier, in the case of Jupiter the surface magnetic field
strength plays a crucial role in the decametric radio emission. Assuming
a magnetic moment of $M_{E,Jup}=1.6\times 10^{30}$~G~cm$^{-3}$, a
planetary radius of $R_{Jup}=7.2\times 10^9$~cm and a dipole field, the Jovian
equatorial magnetic field strength is $\sim 4.3$G, which corresponds to
an electron cyclotron frequency of $\nu_{ce} {(\rm MHz)}=2.8 B=12$~MHz,
which corresponds roughly to the frequency of the peak radio emission
from Jupiter. Based on this we will assume that the peak radio emission
from EGPs will occur at a frequency $\nu_{peak}=M_E/R_p^3$, which we
assume corresponds to the electron cyclotron frequency of the surface
equatorial magnetic field of the EGP.

We do note that this is rather a simplistic assumption and that for
planets in our own solar system there are substantial deviations from a
centred dipole field, with the gas giants having the largest deviations
-- for instance the ratio of maximum to minimum surface magnetic field
should be 2 for a centred dipole, and it is 4.5 for Jupiter and 12 for
Uranus (de Pater \& Lissauer 2001).

The planetary quantities that will affect this frequency will be the
planetary magnetic moment and the planetary radius. We will discuss the
likely values for these two quantities for EGPs below.

\subsubsection{EGP Planetary Magnetic Moment}
\label{sec2p4p1}

The magnetic moment $M_E$ of the EGP is an important parameter, as
$M_E$ is key in setting the size of the planetary magnetosphere
$R_{MP}$, which plays a major role in the expected level of radio
emission and in setting the frequency cut-off of the radio emission
(determined by the maximum magnetic field strength near the planetary
surface). Currently, we have very little information about the
expected magnetic moments for EGPs (see S\'anchez-Lavega 2004), and we
are forced to extrapolate from solar system values.

In Fig.~\ref{fig1} (left panel) we plot the planetary magnetic moments
($M_E$) versus planetary mass for objects in the solar system. This
plot omits the deviant points of Mars and Venus (where we can resort
to arguments about low planetary mass or lack of rotation, though
these do not seem to apply to Mercury, which is perhaps the really
deviant point on this graph). We note a very clear correlation, and
for comparison we also plot a line with $M_E\propto M_p^{1.66}$. Note,
this line is not a fit, merely a schematic representation. If we do a
formal fit for these six points then we find a best-fit slope of
$1.91\pm 0.15$. If we do a fit for the five points, excluding Mercury
(which does seem to be a little discrepant in this diagram), then we
find a best-fit slope of $1.66\pm 0.20$ (which is essentially
identical to that plotted).

We note that for solar system planets (particularly the 4 Jovian
planets), the planetary density is roughly constant, and so that
$M_p\propto R_p^3$, and that this relationship could equally be
$M_E\propto M_p R_p^2$.

Similar related trends have been noted by Arge, Mullan \& Dolginov
(1995), who looked at the correlation between magnetic moment (which
they quantified as $\mu=(B_p R_\ast^3)/2$) and angular momentum ($L=C
M_\ast R_\ast^2 \omega$) for a wide range of bodies (solar system, low
and high mass stellar objects, white dwarfs, pulsars etc). In these
relationships $M_\ast$ and $R_\ast$ are the mass and radius of the
bodies respectively, $B_p$ is the polar field strength at the bodies
surface, $\omega=2\pi/P$ is the rotation frequency and $C$ is a constant
that depends on the mass distribution within the body (and is typically
in the range of $0.1-0.4$ for bodies ranging from normal stars to
compact objects). This relationship was first noted in the context of
the solar system by Blackett (1947), and has been referred to as the
``magnetic Bode's law''.

Arge \etal\ (1995) noted a very clear correlation between $\mu$ and $L$
for the solar system objects and the stellar objects with $\mu\propto
L^\delta$, with $\delta\sim 1$. Taking this relationship and
transferring into the terminology used here for EGPs, we then have

\begin{equation}
M_E\propto M_p R_p^2 \omega\ .
\end{equation}

For the solar system objects (with the exception of Mercury) the values
of $\omega$ are similar (with rotation periods in the range of
$10-25$hours). Indeed, because the range in magnetic moment for the
solar system objects spans so many orders of magnitude, even the
difference between the angular frequency of Mercury and the fastest
rotating planet (a factor 140 or so) does not change the overall trend.
If we plot $M_E$ versus $M_p R_p^2$ instead of $M_p$ we get a plot with
a slope close to unity (Fig.~\ref{fig1}, right panel). The line plotted
in this diagram is not a fit, but a schematic line with slope unity,
plotted so that it passes through the value for Jupiter. Again, if we
do a formal fit to all six points we find a best-fit slope of $1.03\pm
0.1$, and excluding Mercury (as per the left panel of
Fig.~\ref{fig1}), then we find a best-fit slope of $0.87\pm
0.14$. Both of these fitted slopes are unity (within errors).

For massive EGPs (with $M_p>M_{Jup}$) their radius varies very slowly
with mass, and indeed, in many models, shows a slow decrease with
increasing mass (see Section~\ref{sec2p4p2}). This means that we can
assume that $M_E\propto M_p$, based on the values within the solar
system, but also on the more general analysis of Arge \etal\
(1995). The more massive stellar objects in the sample of Arge \etal\
(1995) show the same basic scaling with mass, but in fact the absolute
values of the magnetic moment are higher than for the solar system
objects.  We might expect this relationship not to hold true for very
close in EGPs, which are tidally locked to their host star (see
Section~\ref{sec4}).

Consequently, we assume that the planetary magnetic moment for EGPs is
proportional to the planetary mass. Currently, from the Doppler
solutions, all we have are values of $M_p\sin i$, with $i$ the orbital
inclination of the planet. As noted by Halbwachs (1987), for a randomly
distributed sample, neglecting selection effects, then the distribution
function of the inclinations $i$ will be $\propto \sin i$.  Trimble
(1974), in the context of stellar binary systems, used a mean value of
$\sin^3 i$ to investigate the mass distribution of the component
stars. As discussed by Halbwachs (1987) and Heacox (1995), this
simplistic approach does have some severe limitations (to say the least)
when used to determine the mass distributions of the component stars,
and in the context of EGPs much better statistical methods are available
(for example, Jorissen, Mayor \& Udry 2001). However, for the purposes
required for this paper, where we want to estimate values for individual
planets, this simplistic method is still not an unreasonable
approach. So, in order to have an estimate of the real masses for all
the EGPs listed in Table~2, we simply assume a mean value of the
inclination, such that $<\sin i>=0.866$, corresponding to an inclination 
of $60^\circ$. We note that we could
perhaps improve this flawed approached by adopting $<\sin^3 i>=0.589$ or
$<\sin^3 i>=0.679$ (see Halbwachs 1987), but these different assumptions
gives mean values of $\sin i$ only a few percent different, and not
enough to make a difference to the conclusions of this paper, and the
value we have assumed lies neatly between them

We note that with this value of $i$ some of the planets are relatively
low-mass, with masses as low as $0.23 M_{Jup}$. Here the assumption of
$R_p=R_{Jup}$ may also be a slight overestimate, though we note that
Saturn, even with a mass of $0.3M_{Jup}$, has a radius of $0.84R_{Jup}$.
Leaving this aside, with this assumption we can then determine an
estimate of the planetary mass, and hence the planetary magnetic moment
(see Table~2).  Some additional comments about individual planetary
systems are made in Section~\ref{sec3p1}.

As a final comment, very recently S\'anchez-Lavega (2004) has
published a more detailed physical model of the expected magnetic
moments of EGPs. This analysis supports the notion that EGPs will have
strong magnetic moments, and justifies the scaling of the planetary
magnetic moment with planetary mass, that is seen empirically.
Further, this model includes the effects of planets with a range of
different rotation periods.  S\'anchez-Lavega (2004) concludes that
young, massive and rapidly rotating EGPs are likely to have surface
fields in the range of $30-60$~G, while for the older planets (or
those that are in short-period orbits and are orbitally synchronised)
the surface magnetic fields will be $\sim 1$~G.

\begin{figure}
\vspace*{6.5cm}
\includegraphics{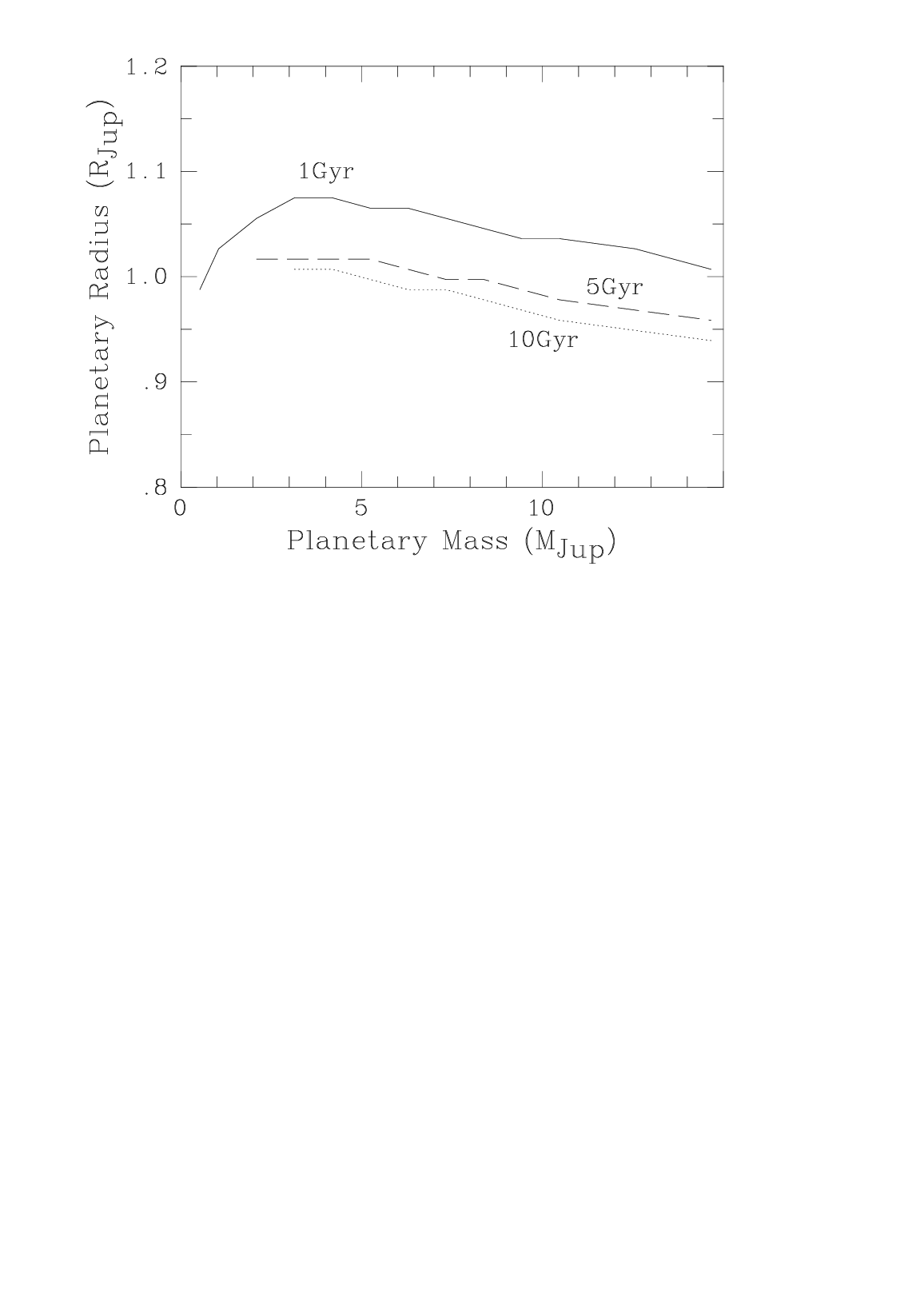} 
\caption{The planetary radius $R_p$ plotted against the planetary mass
$M_p$ for a range of EGP models from Baraffe \etal\ (2003). The plotted
points are for models with ages of 1~Gyr (solid line), 5~Gyr (dashed line)
and 10~Gyr (dotted line).}
\label{fig2}
\end{figure}

\subsubsection{EGP Planetary Radius}
\label{sec2p4p2}

The EGP radius will be an important quantity in the radio emission
from EGPs, entering in to the consideration of the expected frequency
of the peak radio emission.

There is currently very little hard observational information on the
radii of EGPs, and none for planets which are not ``hot Jupiters''.  One
example where there is data is HD\,209458, where the determined radius
is $\sim 1.4R_{Jup}$ (for a planetary mass of $\sim 0.6M_{Jup}$, see
Cody \& Sasselov 2002 and Burrows, Sudarsky \& Hubbard 2003). However,
this planet is a very short period planet (a Class~V ``Roaster'' in the
terminology of Sudarsky \etal\ 2003). For the three currently known
transiting OGLE planets (OGLE-TR-56, 113 and 132), the respective masses
and radii are $0.9 M_{Jup}$ and $1.3 R_{Jup}$ (OGLE-TR-56), $1.4
M_{Jup}$ and $1.1 R_{Jup}$ (OGLE-TR-113), $1.0 M_{Jup}$ and $1.2
R_{Jup}$ (OGLE-TR-132, Bouchy \etal\ 2004). All 4 planets have very
short period orbits and are strongly heated and their radii will almost
certainly be greater than that of a more distantly orbiting EGP.

Zapolsky \& Salpeter (1969) discussed the mass-radius relationship for
cold spheres for masses appropriate for planetary systems and pointed
out that there are substantial differences in the derived radii
depending if the object is hydrogen rich or has a metallic core. They
also showed that the mass-radius relationship has a peak at a mass
around that of Jupiter.

There have subsequently been a large number of more detailed
calculations of the radius of EGPs, and we shall adopt the results of
Baraffe \etal\ (2003) to provide estimates of the EGP radius.  In
Fig.~\ref{fig2} we plot results from the models of Baraffe \etal\
(2003) for EGP models with ages of 1, 5 and 10~Gyr. What we see is
that for a wide range of masses the radii of the EGPs are close to
that of Jupiter (within 10 per cent or so). Given the intrinsic
uncertainties in our knowledge of the actual masses of the EGPs, we
shall (for simplicity) adopt a radius of $R_{Jup}$ for all the EGPs
considered. Most of the planets listed in Table~2 are not extremely
close objects, but some of the important ones are: for example, the
shortest period planet is Tau~Boo\,b (3.3 days), and here we might
expect this planet to have a radius larger than that assumed here
(orbital radius of 0.05~AU).

The assumption that EGPs in the mass range $0.5-15M_{Jup}$ have a
constant planetary radius means that the surface magnetic field rises
linearly with planetary mass $M_p$ (as magnetic moment scales with
planetary mass), and so that the likely frequency of peak emission
$\nu_{peak}\propto M_p$ (which in turn is an important observational
consideration).

\begin{table*}
\caption{The details of the host stars of nearby extrasolar giant
planets considered as possible sources of radio emission. These stars all
lie within 20~pc of the Earth, with the exception of HD\,179949, which is 
included because of the indirect evidence of a magnetosphere. 
The details of the stars are taken from the California and Carnegie
Planet Search website (http://exoplanets.org), as
well as from Santos \etal\ (2003, 2004). The X-ray luminosity data are
from the NEXXUS database (see text for details) and the X-ray flux is
simply $F_X=L_X/(4\pi R_\ast^2)$, and the stellar mass-loss rate is
derived using eqn.~(15). For further details see the text.} 
\begin{tabular}{llcccccc}\hline
Name & Type & Distance & Mass & Radius & X-ray Lum. & X-ray flux & Mass-loss 
rate\\
  & & $D$ & $M_\ast$ & $R_\ast$ & $\log L_X$ & $F_X$ & $\dot M_\ast$ \\
  & & (pc)& (\msun) & ($R_\odot$)& (\ergs) &(erg cm$^{-2}$
s$^{-1}$) & ($\dot M_\odot$) \\ 
\hline

Eps Eri    & K2V & 3.50 & 0.80 & 0.79 & 28.33 &$5.6\times 10^{5}$ & 17.4 \\

Gliese~876 & M4  & 4.69 & 0.32 & 0.41 & 26.49 &$3.0\times 10^{4}$ & 0.16\\

Gliese~86  & K1V & 10.9 & 0.70 & 0.79 & 28.00 & $2.6\times 10^{5}$& 7.2 \\

HD\,3651   & K0V & 11.2 & 0.79 & 0.93 & 27.21 &$3.1\times 10^{4}$ & 0.86\\

55 Cnc     & G8V & 12.5 & 0.88 & 0.93 & -- & -- & --\\

HD\,147513 & G3/G5V & 12.9 & 1.11 & 0.96 &29.03 &$1.9\times 10^{6}$ & 104.7\\

Ups And    & F8V & 13.5 & 1.29 & 1.64 & 28.25 & $1.1\times 10^{5}$ & 11.5\\

47 UMa     & G0V & 14.1 & 1.05 & 1.16 &  -- & -- & --\\

HD\,160691 & G3IV/V & 15.3 & 1.10 & 1.29 & 27.44 &$2.7\times 10^{4}$ & 1.42\\

51 Peg     & G2.5IVa & 15.4 & 1.04 & 1.11 & 26.80 & $8.4\times 10^{3}$ &
0.27\\

Tau Boo    & F7V & 15.6 & 1.30 & 1.44 & 28.99 &$7.7\times 10^{5}$ & 83.4\\

Gliese~777A & G6IV & 15.9 & 0.96 & 1.15 & -- & --& --\\

HD\,128311 & K0V & 16.6 & 0.76 & 0.73  & 28.47  &$9.1\times 10^{5}$ & 26.0\\

HD\,17051  & G0V   & 17.2 & 1.32  & 1.09 & 28.78 & $8.3\times 10^{5}$& 52.1\\

Rho CrB    & G2V & 17.4 & 0.95 & 1.30 & -- & -- & -- \\

HD\,1237   & G6V & 17.6 & 0.99 & 0.82 & 28.94 & $2.1\times 10^6$ & 85.7\\

70 Vir     & G5V  & 18.1 & 0.92 & 1.89 & 27.05 &$5.2\times 10^{3}$ &
0.46 \\

HD\,145675 & K0V & 18.1 & 0.90 & 0.99 & -- & -- & --\\

HD\,39091  & G1V & 18.2 & 1.10 & 1.12 & 27.48 &$4.0\times 10^{4}$ & 1.68\\

HD\,27442  & K2IV & 18.2 & 0.83 & 3.75 & 27.48 &$3.5\times 10^{3}$ & 1.14\\

HD\,217107 & G8IV &19.7  & 0.98 & 1.12 & -- &   -- \\

HD\,192263 & K0V & 19.9 & 0.75 & 0.65 & 27.91 &$3.2\times 10^{5}$ &
6.19\\ \hline

HD\,179949 & F8V & 27.0 & 1.28 & 1.14 & 28.61 &$5.2\times 10^{5}$ & 33.1
\\ \hline 
\end{tabular}
\end{table*}

\begin{table*} 

\caption{The details of the extrasolar giant planets
under consideration as possible sources of radio emission. The EGPs are
ordered according to the distance from the Earth ($D$), and the details
of the planetary mass ($M_p\sin i$), orbital semi-major axis of the
planet $a$ and eccentricity ($e$) are from the California and Carnegie
Planet Search website (http://exoplanets.org). Details of how the
planetary magnetic moment, magnetospheric radius and expected radio flux
are calaculated are described in the text, but the values quoted here
are based on the assumption that $\sin i =0.866$. Where no values of the
magnetospheric parameters are listed it is because the star does not
have an X-ray detection. This means that the star is likely X-ray faint
and consequently unlikely to have a strong wind.}

\begin{tabular}{lcccccccc}\hline 
Name & Period & Planet Mass & Semi-Major & Eccentricity & Magnetic &
Magnetospheric & Radio & Peak\\ & 
&& Axis && Moment & Radius & Flux & Frequency\\ 
&&$M_p \sin i$ &$a$& $e$ & $M_E$ & $R_{MP}$ & $P_{rad}$&$\nu_{peak}$\\
& (days) &($M_{Jup}$) & (AU) && ($M_{E,Jup}$) & ($R_{Jup}$) & (mJy)& (MHz)\\
\hline

Eps Eri\,b & 2548.7 & 0.92 & 3.40 & 0.43 & 1.06 & 25.3 & 2.4 & 10.6\\

Gliese~876\,c & 30.12  & 0.56 & 0.13 & 0.27 & 0.65 & 13.4 & 3.2 & 6.5\\
Gliese~876\,b & 61.02  & 1.89 & 0.21 & 0.10 & 2.18 & 35.3 & 3.8 & 21.8\\

Gliese~86\,b  & 15.77  & 4.01 & 0.11 & 0.04 & 4.63 & 24.9 & 35.1 & 46.3\\

HD\,3651\,b & 62.23 & 0.20 & 0.28 & 0.63 & 0.23 & 6.6 & 0.3 & 2.31\\

55 Cnc\,b  & 14.65  & 0.84 & 0.12 & 0.02 & -- & -- & -- & --\\
55 Cnc\,c  & 44.28  & 0.21 & 0.24  & 0.34 & -- & -- & -- & --\\
55 Cnc\,d  & 5360.0 & 4.05 & 5.9   & 0.16 & -- & -- & -- & --\\

HD\,147513\,b &540.4 & 1.00 & 1.32   & 0.26 & 1.15 & 14.5 & 2.1 & 11.6\\

Ups And\,b  & 4.62  & 0.69 & 0.06 & 0.012 & 0.80 & 5.8 & 21.7 & 8.0\\
Ups And\,c  & 241.5 & 1.89 & 0.83 & 0.28  & 2.18 & 27.3 & 1.3 & 21.8\\
Ups And\,d & 1284.0 & 3.75 & 2.53  & 0.27  & 4.33 & 62.6 & 0.5 & 43.3\\

47 UMa\,b & 1089 & 2.54 & 2.09  & 0.06 & -- & -- & -- & --\\
47 UMa\,c & 2594 & 0.76 & 3.73  & 0.10  & -- & -- & -- & --\\

HD\,160691\,b & 664.2& 1.87 & 1.53 & 0.26 & 2.16 & 47.2 & 0.1 & 21.6\\

51~Peg  & 4.23 & 0.46 & 0.05 & 0.01 & 0.53 & 7.8 & 1.33 & 5.3\\

Tau Boo\,b& 3.3 & 4.13 & 0.05 & 0.01 & 4.77 & 13.0 & 256.0 & 47.7\\

Gliese~777A\,b& 3902  & 1.33 & 4.80 & 0.48 & -- & -- & -- & --\\

HD\,128311\,b& 420.5  & 2.58 & 1.02 & 0.30 & 2.98 & 31.5 & 1.4 & 29.8\\

HD\,17051\,b& 311.3  & 1.94 & 0.91 & 0.24 & 2.24 & 22.3 & 1.9 & 22.4\\

Rho CrB\,b& 39.845 & 1.04 & 0.22 & 0.04 & -- & -- & -- & --\\

HD\,1237  & 133.8 & 3.45 & 0.51 & 0.51 & 3.98 & 24.8 & 8.3 & 39.8 \\

70 Vir\,b& 116.7 & 7.44 & 0.48 & 0.40 & 8.59 & 97.1 & 0.4 & 85.9\\

HD\,145675\,b& 1773.1 & 4.89 & 2.85 & 0.38 & -- & -- & --& --\\

HD\,39091\,b& 2063.8 & 10.35 & 3.29 & 0.62 & 11.95 & 185.2 & 0.1 & 119.5\\

HD\,27442\,b& 423.8 & 1.28 & 1.18 & 0.07 & 1.48 & 34.8 & 0.1 & 14.8\\

HD\,217107 & 7.13 & 1.25 & 0.07 & 0.13 & -- & -- & -- & -- \\

HD\,192263\,b& 24.33 & 0.62 & 0.15 & 0.04 & 0.72 & 8.2 & 1.8 & 7.2 \\\hline

HD\,179949\,b& 3.09 & 0.98 & 0.04 & 0.00 & 1.13 & 5.4 & 23.8 & 11.3 \\\hline
\end{tabular}
\end{table*}

\subsection{The Host Stars of EGPs}
\label{sec2p5}

Extrasolar planets have been, and are expected to be, discovered
around a wide range of stellar types. The initial discovery, of
extrasolar planets orbiting a pulsar PSR~B1257+12, remains perhaps the
most unexpected (Wolszczan \& Frail 1992; see Konacki \& Wolszczan
2003 for a more recent analysis of the results). We note that
Sigurdsson \etal\ (2003) have recently announced the discovery of a
planet mass object and a white dwarf orbiting the pulsar PSR~B1620-26.

Since this initial discovery, extrasolar planets have been discovered
around a range of stars of stellar type F, G, K and M, and the vast
majority of which are main-sequence objects. However, there are now
detections of EGPs around giant stars, such as Iota Draconis (K2III;
Frink \etal\ 2002) and HD\,104985 (G9III; Sato \etal\ 2003), and there
are a number of stars around sub-giants.

In addition to these normal stars it is expected that white dwarf
stars should also host EGPs, and several techniques for finding such
EGPs have been proposed, both in the optical (Burleigh, Clarke \&
Hodgkin 2002; Chu \etal\ 2001) and in the radio (Willes \& Wu
2004). To date no such planet has been found, but it seems likely that
they will exist.

The stellar types of the host stars of EGPs under consideration here
also range from F to M type stars.  For the stellar masses and radii for
the host stars of the EGPs listed in Table~1, we use a variety of
sources, including Santos \etal\ (2003), Santos, Israelian \& Mayor
(2004), and the stellar radii are determined using values for $T_{eff}$
from these sources, and bolometric corrections calculated from Flower
(1996, see also Gaidos 1998), and the Barnes \& Evans relationship (see
Barnes, Evans \& Moffett 1978).

\subsubsection{Stellar Corona and Stellar Winds}
\label{sec2p5p1}

From our own solar system, there appears to be an intimate connection
between the characteristics of the the solar wind and magnetospheric
planetary radio emission. From this we would then expect a similarly
strong connection between the stellar wind characteristics of the host
star and the EGP radio emission. In order to predict the radio
properties of EGPs we need to understand the mass-loss properties of
the host stars.

Stellar winds from stars come with a wide range of characteristics (and
driving mechanisms):

\begin{enumerate}
\item Fast ($1000-5000\kms$) radiatively driven 
winds from early-type stars,
\item Slow (tens of \kms), massive winds from red giants and
supergiants, AGB stars etc, that are driven by pulsations and radiative 
driving of dust,
\item Moderately fast ($\sim 400\kms$)  but diffuse winds from
lower mass coronally active stars, such as the Sun.
\end{enumerate}

While, in principle, the analysis presented here is applicable to any
stellar wind, the host stars of the EGPs under consideration will have
coronally driven winds and we shall concern ourselves with these winds.

Because of the extremely low mass-loss rates, it has hitherto been very
difficult to estimate the mass-loss rates of solar-type stars. This is
in contrast to the OB stars where radio and UV observations can give
good measures of both the mass-loss rate and wind velocity (see
Dougherty \etal\ 2003).

For the solar-type stars, the dominant wind driving mechanism is
associated with a hot corona, as in the Sun. There is a wide range of
coronal properties, with stellar youth leading to much higher levels of
X-ray emission. Naively, one would expected the mass-loss from such
stars to scale, in some way, with coronal X-ray emission (which in turn
is a good indicator of coronal activity).  Recently, Wood \etal\ (2003)
have quantified the connection between stellar X-ray emission and the
expected mass-loss rate, through high resolution Hubble Space Telescope
Lyman\,$\alpha$ observations of nearby coronal stars.

The observed X-ray emission from single stars can be understood in terms
of two different origins, depending on the mass of the star. For high
mass stars, the X-ray emission is believed to be associated with shocks
within the radiatively driven stellar winds of massive stars (Berghofer,
Schmitt \& Cassinelli 1996) while for lower mass stars, relevant to EGPs
here, the X-ray emission is believed to come from a hot corona, analogous
to the situation seen in the Sun. The first mechanism is applicable to O
and early B-type stars, and it is unclear what the origin for the X-ray
emission seen from late B-type through to early F-type stars is, though
it could well often be related to binarity.

The outer convection zone of stars disappears in stars earlier than
spectral type F5, and this is argued to be the reason for the drop of
coronal activity around this spectral range (Stauffer \etal\ 1994), and
so the coronal mechanism for X-rays is applicable for late F stars,
through G and K-stars down to M-stars. The earliest spectral type in our
sample is F7V (Tau~Boo) and so all of our stars will be coronally
emitting.

It is worth noting that the X-ray properties of coronally emitting stars
depend very sensitively on rotation, which in turn depends on age, with
younger stars typically being much more rapid rotators and much more
X-ray luminous than older stars (though there may well be some
exceptions, possibly in the F-stars -- see Suchkov, Makarov \& Voges
2003).

We parameterise the mass-loss rates of the stars in the sample using the
results of Wood \etal\ (2003), and we estimate the total mass-loss rate
from the star ($\dot M_\ast$) from

\begin{equation}
{\dot M_\ast}({\dot M_\odot})= \left[\frac{R_\ast}{R_\odot}\right]^2
\left[\frac{F_X}{F_{X,\odot}}\right]^{1.15\pm 0.20}  
\end{equation}
with $\dot M_\odot$ the mass-loss rate of the Sun ($2\times
10^{-14}\msunyr$) and $F_{X,\odot}$
the solar X-ray surface flux ($F_{X,\odot}=3.1\times
10^4$~erg~cm$^{-2}$~s$^{-1}$). Note that in Wood \etal\ (2003), 
the quoted mass-loss rates are sometimes the mass-loss rates per
unit area, whereas here we always use the total mass-loss rate of the star,
integrated over the entire surface.

We have used the NEXXUS database to
investigate the host star X-ray properties (Schmitt \& Liefke 2004).
The values of the X-ray luminosity and surface
flux for the host stars are shown in Table~1. The X-ray luminosities
range over 2 orders of magnitude from $10^{27}$ to $10^{29}\ergs$, with
an even greater range in surface flux. 
Using the relationship given
above we can now easily determine the overall mass-loss rates for those
stars detected at X-rays, and these too are given in Table~1. We see a
large range in mass-loss rates, from $0.16\dot M_\odot$ for Gliese~876,
up to $>100 \dot M_\odot$ for HD\,147513, with a number of stars
undergoing significantly stronger mass-loss than the Sun.

The X-ray non-detections in this table probably mean that the stars are
relatively faint X-ray sources and are thus unlikely to have winds
substantially stronger than that of the Sun. This in turn means that
they are unlikely to harbour radio bright EGPs.

As discussed in Section~\ref{sec2p3}, the expected radio emission from
EGPs scales as $\dot M_\ast^{2/3}$, and so those stars with the higher
mass-loss rates provide potentially very interesting targets.

Another issues concerns the velocity of the winds, with the radio flux
emitted by an EGP being $\propto V_w^{5/3}$. Unfortunately, we do not
have any information about the expected wind velocities of these stars, and
we resort to assuming a velocity of $V_w=400\kms$ for all stars
in the sample (though see Wood \etal\ 2003 for a discussion of this
assumption).

We note that Cuntz, Saar \& Musielak (2000) used an earlier version of
an X-ray/mass-loss relationship to estimate the energy flux due to the
magnetic interaction of the stellar wind with the EGP.

\section{EGP Magnetospheric Emission: Results}
\label{sec3}

We have collected together all of the necessary data to estimate the
expected radio fluxes from nearby EGPs, and these are listed in
Table~2. What we can see is that while many of the EGPs have very low
levels of expected radio emission, there are a number with levels
greater than 5~mJy. These are (in order of expected levels of emission),
Tau~Boo, Gliese~86, HD\,179949, Ups~And and HD\,1237, with Tau~Boo far
and away the brightest.

We also predict that HD\,179949 should be a (relatively)
bright radio source (and hence to have strong magnetospheric emission),
given the indirect detection of a magnetosphere (Shkolnik
\etal\ 2003). However, it should also be noted that Shkolnik \etal\
(2003) also observed Tau~Boo in the same manner and found no comparable
effect (and we predict that system to be a brighter radio source).

If we consider the Tau~Boo system, then from our analysis 
there are 4 factors that make it so bright:

\begin{enumerate}
\item the proximity $(D=15.6$~pc) of the star to Earth,
\item the orbital distance of the planet from the star ($A=0.05$~AU),
\item the mass of the planet ($M_p\sin i =4.1 M_{Jup}$), and the
corresponding  magnetic moment, and
\item the high mass-loss rate of the star ($\dot M_\ast=83\dot M_\odot$).
\end{enumerate}

Given that $P_r \propto {\dot M}_\ast^{2/3} V_w^{5/3} M_E^{2/3}
A^{-4/3}$, for the case of  Tau~Boo\,b we find that,
compared to Jupiter, the closer orbital distance of the planet
contributes a factor of 450 in increased radio flux, the increased
mass-loss rate a factor 19 and the higher planetary mass, a factor
4. This leads to Tau~Boo\,b being nearly a factor $40000$
times brighter than Jupiter, but nearly $10^6$ times more distant.

The important point is that while the enhanced mass-loss from Tau~Boo
makes a major difference, the dominant reason that Tau~Boo\,b is
expected to be a bright radio source is its orbital distance
(0.05~AU). Indeed, all bar one of the five brightest EGPs are in short
period orbits. The exception is HD\,1237\,b, which has a period of 134
days, and a semi-major axis of 0.5~AU (and an eccentric orbit). The next
longest period of the brightest planets is Gliese~86\,b, with a period
of 16 days (and a semi-major axis of 0.11~AU). However, all of the bright
EGPs do have host stars with winds substantially stronger than the Sun,
and so that to generate a detectable radio flux a short period massive
planet, orbiting a coronally active star is required. This is an
important point and will enable the identification of possible radio
EGPs, using known properties from the orbital solution and X-ray data.

We note that the expected flux from Tau~Boo\,b is in conflict with the
quoted value for a {\it VLA} 74~MHz observation (120~mJy, Farrell \etal\
2003). One problem with the short period EGPs is that they violate at
least one assumption used in the derivation of the radio flux, namely of
a wind velocity of $400\kms$. These planets orbit so close to the
host star (to within  $10R_\ast$ or so), that wind will not have reached 
terminal velocity. Because the radio emission depends on the incident
ram-pressure flux this will have a big impact (in addition, the
magnetospheric structure of close in EGPs will be very different from
that of the Jovian planets in the solar system -- Ip \etal\
2004). Because the radio flux scales as $V^{5/3}$, a reduction in velocity
by a factor 2 results in a radio flux lower by a factor 3.

As discussed by Zucker \& Mazeh (2002) and by Udry, Mayor \& Santos
(2003) there is now evidence that there is a deficit in the number of
high mass planets (with $M_p\sin i > 2M_{Jup}$) with short orbital
periods ($P_{orb}<100$~days). This is the case for planets orbiting
single stars (which makes up the bulk of the currently known sample),
but the opposite (i.e. an excess of high mass planets in short period
orbits) seems to be true for planets in binary stellar systems, pointing
towards different formation/migration mechanisms (though the number of
examples in this second category is small). As discussed here, massive
planets in short period orbits is exactly the situation required for
significant radio emission, and so, to some extent, is selected against
by nature. This will have implications for the use of radio surveys in
detected EGPs.

It is worth noting that of the expected radio brightest systems, both
Tau~Boo and Gliese~86 are binary systems, and of the other
stars in our sample, Eps~Eri is also a reported binary.

Another consideration is the effect of plasma frequency in those systems
with short period planets and high mass-loss rates. From eqn.~(1) we have

\begin{equation} 
\nu_{pe}({\rm MHz})=9\times 10^{-3} n_e^{1/2} =
0.03\left(\frac{\dot M_\ast}{V_{400}A^2}\right)^{1/2} 
\end{equation}
with $A$ in AU, $\dot M_\ast$ in units of $\dot M_\odot$ and the wind
velocity in units of $400\kms$. For a star with a high mass-loss rate,
$\dot M_\ast=100\dot M_\odot$, $A=0.05$~AU, and $V=200\kms$ the plasma
frequency in the stellar wind at the radius of the planet is 8.5~MHz, a
frequency which will impinge on the radio emission, particularly of lower
mass planets.

These parameters are close to what is appropriate for Tau Boo, and
although this plasma frequency is lower than the expected peak radio
emission ($\sim 50$~MHz), it is reasonably close and means that for some
lines of sight that pass even closer to the star there may be a problem
with the planetary radio emission being screened out (depending on
system inclination), and so care should be taken when observing
Tau~Boo to avoid the possibility of plasma frequency screening, and
epochs when the planet is in front of the star are to be preferred.

\subsection{Comments on Individual Systems}
\label{sec3p1}

For a few systems, there is some additional information about the orbital 
solution (specifically the inclination) for the EGP.

In the case of Gliese~876, recent {\sl Hubble} Space Telescope
astrometric observations, reported on by Benedict \etal\ (2002), have
constrained the orbital inclination of the system (or more accurately,
the most massive planet Gliese~876\,b). The derived value of the
inclination is $i=84^\circ \pm 6^\circ$. This means that the mass of
Gliese~876\,b is $M_p=1.89 M_{Jup}$, and the value for the planetary
magnetic moment quoted in Table~2 is overestimated by $15\%$. As $P_r
\propto M_E^{2/3}$, the quoted expected radio flux for Gliese~876\,b is
overestimated by $\sim 10\%$. However, Gliese~876\,b is not one of the
brightest targets.

In the case of Eps~Eri, Gatewood (2000) report on astrometric
observations, and derive an inclination of $i=46^\circ \pm 17^\circ$,
with corresponds to $\sin i = (0.72^{+0.17}_{-0.24})$, and a
planetary mass of $M_p=1.28 M_{Jup}$. This means that the values quoted
for the magnetic moment for Eps~Eri\,b is underestimated by $\sim 20\%$,
and the expected radio flux underestimated by $\sim 10\%$. Again,
Eps~Eri\,b is not one of the expected brightest radio sources.

\subsection{Eccentric orbits}
\label{sec3p2}

We have calculated the radio flux assuming that the planet is orbiting
at the distance of the semi-major axis axis of the orbit
(i.e. $A=a$). For some of the planets with large eccentricities, this
will be only a rough average value. We have seen that we expect the
radio flux to scale as $A^{-4/3}$, and so, for example, for HD\,3651\,b,
with an eccentricity of $e=0.63$, the planet will be a factor 3.8
brighter than the value quoted in Table~2 at periastron, and a factor
0.5 fainter at apastron. For other planets the values are
correspondingly smaller. For the radio bright EGPs (with the exception
of HD\,1237), which are short period planets, the eccentricities are low
($e=0.04$ at most).

The planet orbiting HD\,1237 is the most interesting example, with an
eccentricity of $e=0.51$. This means that although the mean flux
expected is 8.3~mJy, we would expect this to range from 4.8 to 21.5~mJy,
making HD\,1237 a potentially very interesting source if observed around
periastron. However, the location on the sky of HD\,1237
($\alpha(J2000)=00^h 16^m 12.7^s$, $\delta(J2000)= -79^\circ 51' 04''$)
means that it is not convenient for observations from either the {\it
VLA} or {\it LOFAR} (if sited in Holland).

Consequently, the influence of orbital eccentricity on expected radio
emission is not predicted to be major for most of the candidate
systems. However, in certain systems, such as HD\,1237, it is important
and should be noted when choosing to observe this system.

\section{Discussion and Conclusions}
\label{sec4}

In this paper we have presented a critical discussion of the expected
radio properties of EGPs in the light of both the planetary and stellar
characteristics of the host star. As a result of this we can make the
following statements about the likely levels of EGP radio emission.

\begin{enumerate}

\item Only very nearby EGPs are likely to be detected by their radio
emission. The $1/D^2$ fall-off makes the problem very hard. There are
only 28 EGPs currently known within 20~pc (orbiting 22 different host
stars) and only 3 within 10~pc (one orbiting Eps Eri and two orbiting
Gliese~876).

\item The prime factor in determining whether an EGP is radio bright is
its orbital period. Shorter period planets will also show more radio
emission, with the radio flux scaling as $A^{-4/3}$. However, this
effect will not continue indefinitely as if the planet is very close to
the star the incident wind will impinge on the magnetosphere at lower
velocities. Also, eventually plasma-frequency screening will begin to
play a role for very short period planets.

\item Higher levels of radio emission are likely to be associated with
more massive planets, with the radio flux scaling as $M_p^{2/3}$. The
frequency of the radio emission from the more massive planets will
likely be shifted to more amenable wavelengths. We note that the most
massive planet within 20~pc is HD\,39091\,b (with $M_p\sin
i=10.35M_{Jup}$), though we do not expect this planet to be a bright
radio source, as it is in a very long period orbit.

\item Higher levels of EGP radio emission are expected from stars with
higher coronal X-ray surface fluxes (and hence higher mass-loss
rates). Of the nearby stars (within 20~pc) a fair number have been
detected at X-ray wavelengths and have mass-loss rates in the range of
$\dot M_\ast = 0.1-100 {\dot M_\odot}$. The nearest star harbouring an
exoplanet (Eps Eri) does have a strong wind, but also a long period
planet, which offsets this. Tau~Boo, which remains the most promising
candidate, also has a very strong wind, with $\dot M_\ast=83\dot M_\odot$,
although HD\,1237 is another promising candidate, has $\dot M_\ast=86\dot
M_\odot$.

\item The peak frequency of emission in the five brightest cases range
from $\sim 10$~MHz up to 50~MHz, though these values are
uncertain. Certainly, very low frequency observations will be required
for a detection, with even the 74~MHz {\it VLA} observations possibly
being at too high a frequency. As with Jupiter, we do expect emission
from the EGPs at $\nu>\nu_{peak}$, but it will likely be falling off
very rapidly.

\item Two of the five predicted brightest systems (Tau~Boo and
Gliese~86) are in binary stellar systems (there seems to be a dearth of
massive planets in short period orbits in single stellar systems and the
apparent opposite trend in binary systems, see Zucker \& Mazeh 2001).

\item In some cases, orbital eccentricity may play a major role in
detectability of the system. The prime example from this study is
HD\,1237, where the expected flux will vary by a factor 4.5 over the 134 
day orbit, reaching over 20~mJy.

\end{enumerate}

On the basis of the results presented here, there remains the prospect
of detecting nearby EGPs (particularly with {\it LOFAR} or {\it SKA}).
The best candidates (in order of expected brightness), with their
predicted peak frequencies) are:

\begin{enumerate}
\item Tau Bootes (48~MHz)
\item Gliese 86 (46~MHz)
\item HD\,179949 (11~MHz)
\item Upsilon Andromeda (8~MHz)
\item HD\,1237 (40~MHz)
\end{enumerate}

Previous searches for EGPs (see Section~\ref{sec1p2}) have tended to
focus on rather higher frequencies than these (with the 74~MHz
observations at the {\it VLA} being the most sensitive). The emission from
Jupiter falls off very sharply with increasing frequency, which means
that careful (multi-frequency) observations are needed.

On a positive note, all five of these objects have fluxes that are
potentially detectable, particularly with upcoming {\it LOFAR}
telescope, which will operate at lower frequencies, more appropriate
for detecting EGPs.. The quoted sensitivities for {\it LOFAR} (for a 1
hour integration with the full array, single polarisation, 4~MHz
bandwidth), are 3~mJy at 10~MHz, 1.6~mJy at 30~MHz and 1.0~mJy at
75~MHz (see http://www.lofar.org). However, not all of them are
visible from the anticipated {\it LOFAR} site (HD\,1237 and Gliese~86
are not -- Gliese~86 is just visible from the {\it VLA}). The other 3
targets are visible from this site.

As a further consideration, it is notable that Shkolnik \etal\ (2003)
also detected night-to-night activity on Tau~Bootes and Upsilon
Andromeda, as well as HD\,179949 (though in these cases the
synchronicity with the planet was not obvious). This activity was not
present in the standard stars observed, nor in the 51~Peg system. This
activity could also be related to magnetospheric activity.

There are however, many uncertainties in this analysis, such as the
extrapolations from our own solar system to extrasolar systems, in
particular how the radio flux from short period systems will depend on
the stellar/planetary parameters. In addition, we have assumed a scaling
for the magnetic moment with mass. The close-in planets may well be
rotationally locked to their host star. What this lack of rotation will
do to any planetary dynamo and magnetic field is uncertain, but is
likely to reduce it. If this turns out to be the case then HD\,1237 is
perhaps the most promising target, with a reasonably long period planet
(and likely interesting orbital variability).

As a further consideration, the expected radio bright EGPs will
preferentially be orbiting young, rapidly rotating and X-ray luminous
stars (with high mass-loss rates). This means these planets themselves
will be young too.  Given that it may take planets a reasonable length
of time to generate a self-sustaining magnetic field, it is possible
that this may limit the radio brightness of EGPs, with increasing
magnetic field strength of the EGPs being coupled with decreasing
mass-loss rate of the host star. Further study of the likely temporal
evolution of magnetic fields in EGPs is needed.

We have also assumed that the host stars are not radio emitters at
similar wavelengths, but we do note that radio flare stars may well
confuse the issue in some cases. This may be an issue for some 
M stars, but is much less likely to be an issue for F, G and 
K-stars (Bastian 1990). In addition, while the winds of massive stars
can also generate both thermal and non-thermal radio emission, the winds 
considered here are unlikely to be strong enough to generate any
confusing emission.

In addition to the expected flux, the optimum frequency to observe these
sources is also uncertain.  The predicted peak frequencies are around
50~MHz for Tau~Boo and Gliese~86 (which is still at a lower frequency
than the 74~MHz observations of Tau~Boo), and rather lower for
HD\,179949 and Ups~And.  The peak frequency is, however, very uncertain,
and depends on the planetary magnetic moment (and planetary mass), which
for all of these objects is uncertain due to the $\sin i$ factor (and
the assumptions made in this model). We have assumed a mean value of the
inclination, and it could be for these objects that $i$ is closer to
$90^\circ$, which will reduce the expected flux as well as making the
peak frequency difficult to observe (though the converse is also
possible, making the situation easier). The peak frequency also depends
on the planetary radius, which we may be underestimating for the very
close planets (those with orbital separations of 0.05~AU).  Gliese~86\,b
has an orbital separation of 0.11~AU, putting in the category of a
class~IV EGP, and so, while it is possible we are underestimating the
planetary radius, the effect is likely to be less than for Tau~Boo.

On a positive note, given the anticipated parameters of the {\it LOFAR}
instrument, from the estimates presented here, there should be ample
sensitivity to detect a substantial number of EGPs. These EGPs will have
a range of period/mass characteristics, and thus the observations will
initiate the observational study of extrasolar magnetospheres.

So, in summary, we have performed a detailed analysis of the
expected radio emission from nearby extrasolar giant planets. In
particular, new results allow us for the first time to be quantitative
about the expected stellar wind properties of the host stars of EGPs,
which in turn has important consequences on their radio detectability.
We predict that a number of EGPs will emit at radio wavelengths at
levels that are likely to be observable with {\it LOFAR}. We do note that
there remain a number of issues that could reduce the observability of
these objects, and care should be taken when choosing to observe the
candidates, but nonetheless there does remain the prospect of a
detection and we have proposed the five most promising targets on which
to undertake a search.

\section*{Acknowledgements}

I would like to acknowledge the helpful comments of the anonymous
referee, and also comments made on an earlier version of the paper by Steve
Spreckley.

\end{document}